\documentclass[conference]{IEEEtran}
\IEEEoverridecommandlockouts
\usepackage{cite}
\usepackage{amsmath,amssymb,amsfonts}
\usepackage{algorithmic}
\usepackage{textcomp}
\usepackage{xcolor}
\usepackage[pdftex]{graphicx}
\usepackage{epstopdf}

\usepackage{amsmath,amssymb,enumerate} 
\usepackage{bm} 
\usepackage{indentfirst} 
\usepackage{empheq} 
\usepackage{here} 
\newcommand{\dd}{\mathrm{d}}

\newcommand{\ii}{\mathrm{i}}

\def\BibTeX{{\rm B\kern-.05em{\sc i\kern-.025em b}\kern-.08em
    T\kern-.1667em\lower.7ex\hbox{E}\kern-.125emX}}

\begin{document}

\title{Closed-Form Solutions of the Fundamental Equation That Describes User Dynamics\\ in Online Social Networks
}

\author{\IEEEauthorblockN{T. Ikeya}
\IEEEauthorblockA{\textit{Graduate School of Systems Design} \\
\textit{Tokyo Metropolitan University}\\
Hino, Tokyo 191-0065, Japan \\
{\tt t.ikeya12345@gmail.com}}
\and
\IEEEauthorblockN{Masaki Aida}
\IEEEauthorblockA{\textit{Graduate School of Systems Design} \\
\textit{Tokyo Metropolitan University}\\
Hino, Tokyo 191-0065, Japan\\
{\tt aida@tmu.ac.jp}}
}

\maketitle

\begin{abstract}
The oscillation model, based on the wave equation on networks, can describe user dynamics in online social networks.
The fundamental equation of user dynamics can be introduced into the oscillation model to explicitly describe the causal relation of user dynamics yielded by certain specific network structures.
Moreover, by considering the sparseness of online social networks, a novel fundamental equation of different form has been devised.
In this paper, we derive a closed-form solution of the new fundamental equation.
Also, we find the closed-form solution of the new fundamental equation can generate the general solution of the original wave equation. 
\end{abstract}

\begin{IEEEkeywords}
online social networks, oscillation model, user dynamics, causality, anti-commutation relation
\end{IEEEkeywords}

\section{Introduction}
\label{sec.1}
In recent years, people around the world have been increasingly using social networking services (SNSs).
For example, YouTube and TikTok are used to distribute and watch videos, Instagram and Flickr are used to share images, and Twitter and Weibo are used to share short texts and videos, and images.
Furthermore, information obtained from one SNS can be shared with other SNSs.
With the widespread use of SNSs, people can easily transmit and exchange information in online social networks (OSNs).
Accordingly, SNSs can bring benefits to people in the virtual space to facilitate the enrichment of friendships and collect information about their preferences.

However, the SNSs also have their negative aspects.
Prime examples are the online flaming phenomenon caused by collective user dynamics and the online echo-chamber phenomenon in which specific opinions and beliefs are strengthened among people in a closed community.
These phenomena are likely to negatively impact not only on online communities such as SNSs but also on activities in the real world.
To consider countermeasures to such phenomena, it is necessary to understand the characteristics of user dynamics caused by interactions between users via OSNs.

As one of the models created to describe user dynamics, the previous study \cite{oscillation_model,oscillation_model2} proposed an oscillation model based on the wave equation in order to describe user dynamics in OSNs.
Additionally, in the framework of the oscillation model, the fundamental equations that can explicitly describe the causal relationship between user dynamics and certain specific network structures were introduced \cite{fundamental_equation_1,fundamental_equation_2}.
These fundamental equations of OSNs are of two types: one cannot consider the link structure whereas the other can.
The first type of fundamental equation is easy to solve and we can obtain closed-form solutions, but no closed-form solutions of the second type of fundamental equation have been published so far.
This paper derives a closed-form solution of the second type of fundamental equation for OSNs.
Also, we show the closed-form solution of the second type of fundamental equation can generate the general solution of the original wave equation. 

The rest of this paper is organized as follows.
In Sec.~\ref{sec.2}, we describe studies on SNS and user dynamics related to this work and explain its innovation.
In Sec.~\ref{sec.3}, we overview the oscillation model based on the wave equation on networks and describe two different fundamental equations, they are called boson-type and fermion-type equations. 
The boson-type equation cannot consider the link structure of OSNs, but the fermion-type equation can. 
In Sec.~\ref{sec.4}, we derive a closed-form expression of the fermion-type fundamental equation and show that it differs from the solution of the boson-type fundamental equation. 
In Sec.~\ref{sec.5}, we show the closed-form solutions of the fermion-type fundamental equation can generate a general solution of the original wave equation.
In Sec.~\ref{sec.6}, we present our conclusions.

\section{Related Work}
\label{sec.2}
Explosive user dynamics, including online flaming phenomena, can seriously impact online communities and social activities in the real world. 
Such user dynamics are generated from user interactions and manifest themselves as a divergence of user activity intensity.
Therefore, understanding the user dynamics in OSNs is an important issue.

Studies on user dynamics in OSNs have examined various models in an effort to capture the diversity of the characteristics of user dynamics. 
User dynamics that describe the adoption and abandonment of a particular SNS have been modeled by the SIR model, which is a traditional epidemiological model, and the irSIR model, which is an extension of that model~\cite{SIR,Nekovee2007,Cannarella2014EpidemiologicalMO}. 
These models express the state transition of an objective system in a macroscopic framework but are not good at describing individual user dynamics. 
Also, these models describe the speed of changes in transient states and/or the configuration of the final steady-state of the system, but divergence in the stability of user dynamics is not addressed by these models. 

The consensus problem including user opinion formation is typical of the dynamics in OSNs~\cite{Olfati-Saber2004,Wang2010}. 
This can be modeled by a first-order differential equation with respect to time using a Laplacian matrix that represents the social network structure.
The differential equation used in this model is a sort of continuous-time Markov chain on the network.
First-order differential equations with respect to time are also used in modeling of the temporal change in social network structure (linking or delinking of the nodes), and there are models that consider change via a continuous-time Markov chain~\cite{Snijders2010}. 
In addition to theoretical models, \cite{Cha2009} and \cite{Zhao2012} studied user dynamics analysis based on real network observations. 
Similar to epidemiological models, the Markov chain describes the transient states and/or the steady-state of the system, but not the divergence of user dynamics. 

The oscillation model is introduced for describing user dynamics in OSNs including online flaming phenomena \cite{oscillation_model}.
It is based on the wave equation on networks. 
Although the wave equation may not be familiar in the field of network engineering,  it can describe the propagation of a certain influence between users at a {\em finite speed}.
In this case, the wave equation describes the propagation of influence between users via OSNs. 
The oscillation energy calculated from the oscillation model gives a generalized notion of conventional node centrality (degree centrality and betweenness centrality) \cite{takano,takano2} and the oscillation model can describe explosive user dynamics including online flaming as divergence in the oscillation energy. 
Moreover, the oscillation model yields fundamental equations that can describe not only user dynamics but also causal relationships between user dynamics and the structure of OSNs \cite{fundamental_equation_1,fundamental_equation_2}.

There are two types of fundamental equations: boson-type and fermion-type \cite{ITC2020}. 
The boson-type fundamental equation cannot consider the structure of links, whereas the fermion-type can take account of characteristics such as the sparseness of the link structure of OSNs.
The solutions of both types of fundamental equations can generate solutions of the original wave equation of OSNs. 

For the boson-type fundamental equation, the closed-form solution is easily obtained, but the closed-form solution of the fermion-type equation is not obtained yet.
Since the fermion-type fundamental equation is more suitable for describing user dynamics in actual OSN structures, closed-form solutions of the fermion-type fundamental equation remain the desired goal. 

In this paper, we derive a closed-form solution of the fermion-type fundamental equation.
Also, we show general solutions of the original wave equation generated from the closed-form solution of the fermion-type fundamental equation.

\section{Oscillation Model for Online Social Networks}
\label{sec.3}
According \cite{oscillation_model,fundamental_equation_1}, we briefly explain the oscillation model for describing user dynamics in OSNs, and introduce two types of fundamental equations.

\subsection{Wave Equation-Based Model for Online User Dynamics}
Let $\mathcal{G}(V,\, E)$ be a directed graph representing the structure of an OSN with $n$ nodes, where 
$V=\{ 1, \, 2, \, \dots, \, n \}$ denotes the set of nodes and $E$ denotes the set of links.
For a directed link from node $i$ to node $j$, $(i \rightarrow j) \in E$, the weight of directed link $(i \rightarrow j)$ is denoted by $w_{i \, j}>0$; the adjacency matrix $\bm{\mathcal{A}}:=[\mathcal{A}_{i \, j}]_{1 \leq i, \, j \leq n}$ is defined as
\begin{align*}
  \mathcal{A}_{ij} :=
  \begin{cases}
    w_{ij}, & ((i\rightarrow j) \in E),\\
    0, & ((i\rightarrow j) \notin E).
  \end{cases}
\end{align*}
Furthermore, the weighted nodal degree of node $i$ is defined as $d_{i}:=\sum_{j \in \partial i}w_{ij}$ and the degree matrix is given as $\bm{\mathcal{D}}:={\rm diag}(d_{1}, \, d_{2}, \, \dots, \, d_{n})$,
where $\partial i$ is the set of nodes adjacent to node $i$.
Also, the Laplacian matrix of $\mathcal{G}(V,\, E)$ is defined as $\bm{\mathcal{L}}:=\bm{\mathcal{D}}-\bm{\mathcal{A}}$.

The oscillation model is a minimal model for describing user dynamics, that is, we assume a universal model as simple as possible.
First, assuming that the state of each node can be described by a simple one-dimensional parameter, we let $x_{i}(t)$ be the state of node $i$ at time $t$ and $\bm{x}(t):={}^t\!(x_{1}(t), \, x_{2}(t), \, \dots, \, x_{n}(t))$ be an $n$-dimensional state vector for all nodes.
Next, we introduce the interaction between nodes. 
Between adjacent nodes $i$ and $j$, a force acts in a direction so as to reduce the difference in node state between nodes $i$ and $j$.
The strength of this force is proportional to the absolute value of the difference in state quantities: $|x_{i}(t) - x_{j}(t)|$.
The equation of motion of the user state vector in OSNs is expressed as follows: 
\begin{align}
  \frac{{\rm d}^2}{{\rm d}t^{2}}\, \bm{x}(t) = -\bm{\mathcal{L}} \, \bm{x}(t).
  \label{eom}
\end{align}
This is called the wave equation on networks.
The wave equation (\ref{eom}) describes a phenomenon that inter-user influence propagates via OSNs at a {\em finite speed}.

\subsection{Fundamental Equation of Online User Dynamics}
The fundamental equation is introduced for describing the causal relationship between user dynamics and the structure of OSNs. 
This means that we can understand the effect of a certain specific structure of networks on user dynamics. 
Consider the situation that we completely know the relationship between the structure of an OSN and user dynamics generated from the OSN. 
Here, if we change the structure of the OSN by adding some links to the OSN, the difference of the newly generated user dynamics can be understood by using a newly added graph structure. 
To describe the causal relationship, it is necessary to describe user dynamics by a first-order differential equation with respect to time \cite{oscillation_model,fundamental_equation_1}. 
Since the equation of motion (\ref{eom}) is a second-order differential equation with respect to time, it cannot describe the causal relations. 
The fundamental equation is a first-order differential equation with respect to time and can describe user dynamics and causal relations.

Let us consider OSNs whose Laplacian matrix has only real eigenvalues. 
This constraint is needed because non-real eigenvalues of the Laplacian matrix cause the divergence of oscillation energy as is seen in online flaming. 
We now concentrate on the situations that there the strength of user dynamics does not diverge. 
In this case, the Laplacian matrix is a semi-positive definite matrix, and its square root is uniquely determined as a semi-positive definite matrix. 
Let the square root matrix of $\bm{\mathcal{L}}$ be $\sqrt{\bm{\mathcal{L}}}$, which is an $n\times n$ matrix and $(\sqrt{\bm{\mathcal{L}}})^2 = \bm{\mathcal{L}}$. 
Using $\sqrt{\bm{\mathcal{L}}}$, the fundamental equation is expressed as \begin{align}
  \pm {\rm i} \, \frac{{\rm d}}{{\rm d}t} \, \bm{x}^{\pm}(t)
  =\sqrt{\bm{\mathcal{L}}} \, \bm{x}^{\pm}(t),
  \  \text{(double sign corresponds)},
  \label{fundamental_equation_eom}
\end{align}
where $\bm{x}^{\pm}(t)$ is an $n$-dimensional vector. 
The solutions $\bm{x}^{\pm}(t)$ are also solutions of the original wave equation (\ref{eom}). 

\begin{figure}[bt]
  \centering
    \includegraphics[width=0.95\linewidth]{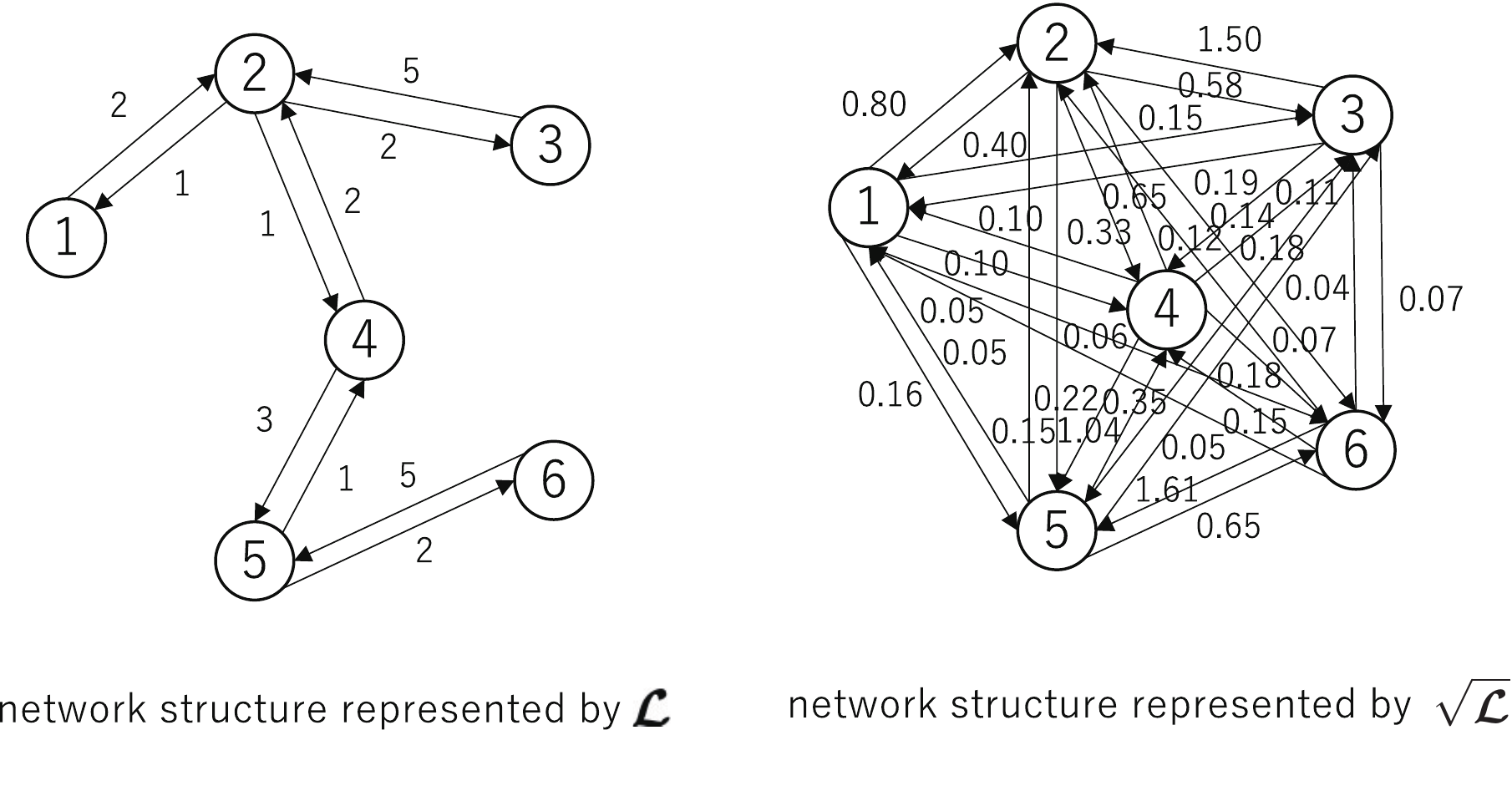}
    \caption{Network structures represented by $\bm{\mathcal{L}}$ and $\sqrt{\bm{\mathcal{L}}}$}
    \label{laplacian_and_root}
\end{figure}
By introducing $2n$-dimensional vector 
\[
\bm{\Hat{x}}(t) := \bm{x}^{+}(t)\otimes
  \begin{pmatrix}
    1\\
    0
  \end{pmatrix}+\bm{x}^{-}(t)\otimes
  \begin{pmatrix}
    0\\
    1
  \end{pmatrix},
\]
the fundamental equation (\ref{fundamental_equation_eom}) is expressed by the following equation
\begin{align}
  {\rm i} \, \frac{{\rm d}}{{\rm d}t}  \, \bm{\hat{x}}(t) = \left(\sqrt{\bm{\mathcal{L}}} \otimes 
  \begin{bmatrix}
  +1 & 0\\
  0 & -1
\end{bmatrix}
\right) \bm{\hat{x}}(t),
 \label{fundamental_equation_eom_two}
\end{align}
where $\otimes$ denotes the Kronecker product \cite{kronecker}.
We call the fundamental equation \eqref{fundamental_equation_eom_two} (and equivalently \eqref{fundamental_equation_eom}) the boson-type fundamental equation. 

The closed-form solution of the fundamental equation~(\ref{fundamental_equation_eom}) is easily obtained as 
\begin{align}
\bm{x}^\pm(t) &= \bm{P} \, \exp(\mp\ii\,\bm{\Omega}\,t) \, \bm{P}^{-1} \, \bm{x}^\pm(0), 
\label{solution-x-pm}\\
&\qquad\qquad \text{(double-sign corresponds)}. 
\notag
\end{align}
where $\bm{P}$ is a regular matrix consisting of the eigenvectors of the Laplacian matrix $\bm{\mathcal{L}}$. 
The diagonal matrix is defined as 
\[
\bm{\Lambda} := \text{diag}(0,\,\lambda_1,\,\dots,\,\lambda_{n-1}),
\]
where$\lambda_\mu$ $(\mu=0,\,1,\,\dots,\,n-1)$ denotes eigenvalues of $\bm{\mathcal{L}}$ and $\lambda_0 = 0$.
$\bm{\mathcal{L}}$ is diagonalized as  
\[
\bm{\Lambda} =  \bm{P}^{-1} \,  \bm{\mathcal{L}} \, \bm{P}. 
\]
In addition, 
\[
\bm{\Omega} := \sqrt{\bm{\Lambda}} = \mathrm{diag}(0,\,\omega_1,\,\dots,\,\omega_{n-1}),
\]
and, for eigenvalue $\lambda_\mu$ of $\bm{\mathcal{L}}$, $\omega_\mu = \sqrt{\lambda_\mu}$. 
Rewriting the above solution as the solution of the fundamental equation~(\ref{fundamental_equation_eom_two}) is expressed as follows:
\begin{align}
\bm{\hat{x}}(t) &= \Big(\bm{P} \, \exp(-\ii\,\bm{\Omega}\,t) \, \bm{P}^{-1} \otimes 
\begin{bmatrix}
+1 & 0\\
0 & 0
\end{bmatrix}
\notag\\
&\quad\quad\quad + \bm{P} \, \exp(+\ii\,\bm{\Omega}\,t) \, \bm{P}^{-1} \otimes 
\begin{bmatrix}
0 & 0\\
0 & +1
\end{bmatrix}
\Big)\, \bm{\Hat{x}}(0) 
\label{solution-sqrt{L}}.
\end{align}

Although the boson-type fundamental equation \eqref{fundamental_equation_eom_two} can describe both user dynamics and the causal relation, they have an unacceptable problem.
The link structure of $\sqrt{\bm{\mathcal{L}}}$ is generally a complete graph even if the network represented by $\bm{\mathcal{L}}$ has a sparse structure.
In OSNs, the situation that all users of the world are connected directly is obviously an unacceptable situation. 
Figure~\ref{laplacian_and_root} shows an example of such a situation. 
The left panel shows an example of OSNs with 6 nodes. 
The square root of the Laplacian matrix shown in the left panel corresponds to a complete graph shown in the right panel. 
This is an unacceptable situation because there are extra links set between unrelated users.
The matrix appearing in the fundamental equation should have exactly the same link structure as the original OSNs 

\begin{figure}[bt]
  \centering
    \includegraphics[width=0.95\linewidth]{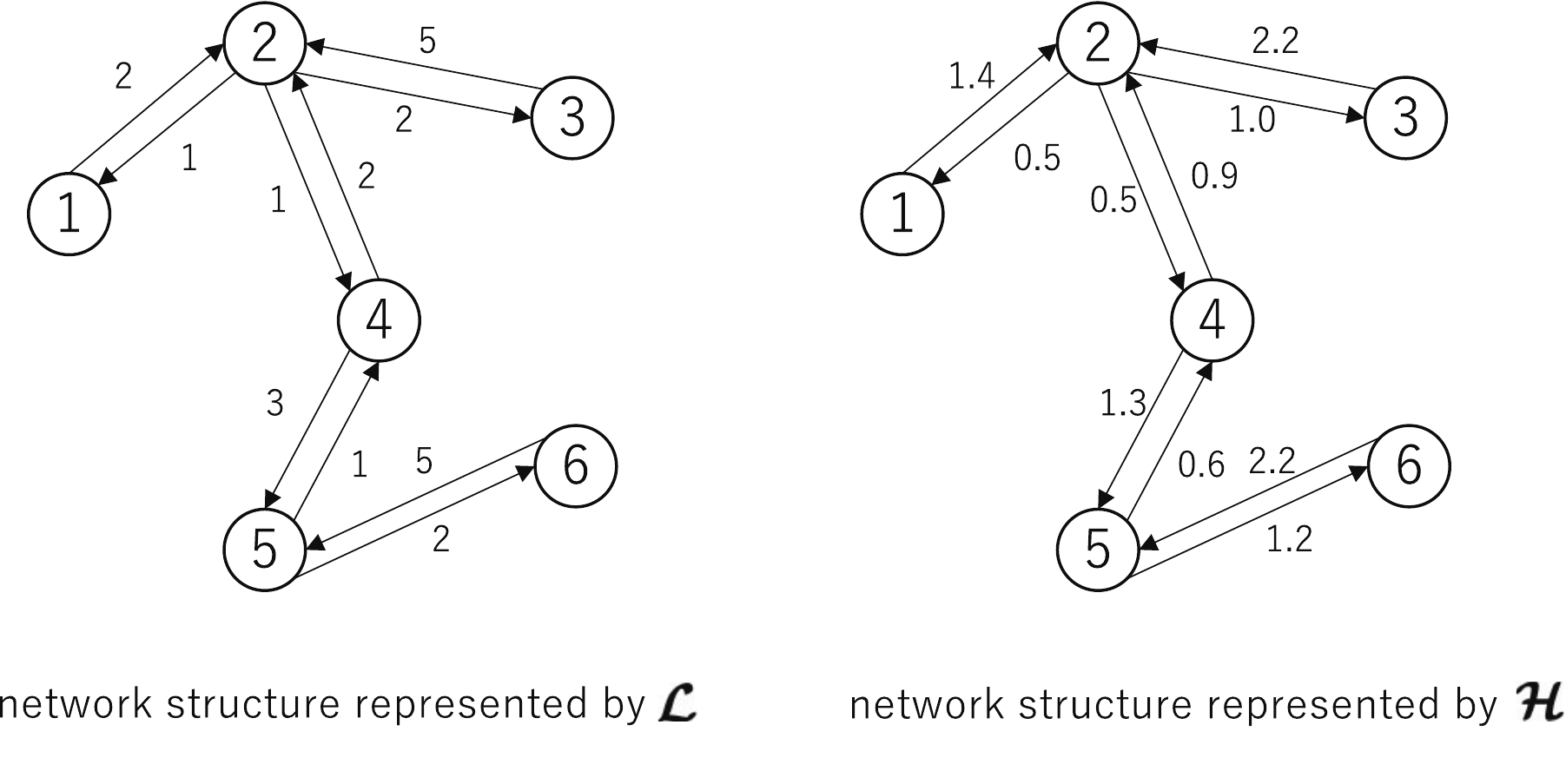}
    \caption{Network structures represented by $\bm{\mathcal{L}}$ and $\bm{\mathcal{H}}$}
    \label{laplacian_and_root_2}
\end{figure}

The fermion-type fundamental equation can avoid the above problem. 
First, we introduce the semi-normalized Laplacian matrix. 
As the well-known normalized Laplacian matrix is defined as
\[
\bm{\mathcal{N}} 
:= (\sqrt{\bm{\mathcal{D}}})^{-1} \, \bm{\mathcal{L}} \, (\sqrt{\bm{\mathcal{D}}})^{-1}
= \bm{I} - (\sqrt{\bm{\mathcal{D}}})^{-1} \, \bm{\mathcal{A}} \, (\sqrt{\bm{\mathcal{D}}})^{-1} ,
\]
let us define the semi-normalized Laplacian matrix as
\[
\bm{\mathcal{H}} 
:= (\sqrt{\bm{\mathcal{D}}})^{-1} \, \bm{\mathcal{L}}
= \sqrt{\bm{\mathcal{D}}} - (\sqrt{\bm{\mathcal{D}}})^{-1} \, \bm{\mathcal{A}},
\]
where $\bm{I}$ is the $n\times n$ unit matrix. 
Note that the semi-normalized Laplacian matrix is a kind of Laplacian matrix whose link weight of $(i\rightarrow j)$ is $w_{ij}/d_i$. 
In addition, the link structure of $\bm{\mathcal{H}}$ is completely the same as that of $\bm{\mathcal{L}}$ with respect to the presence or absence of links (see Fig.~\ref{laplacian_and_root_2}).

Second, let us introduce the $2n\times 2n$ matrix $\bm{\mathcal{\Hat{H}}}$ defined as 
\begin{align}
  \bm{\mathcal{\hat{H}}}:=\sqrt{\bm{\mathcal{D}}}\otimes
  \begin{bmatrix}
    +1 & 0\\
    0 & -1
  \end{bmatrix}
  -(\sqrt{\bm{\mathcal{D}}})^{-1} \, \bm{\mathcal{A}}\otimes \frac{1}{2}
  \begin{bmatrix}
    +1 & +1\\
    -1 & -1
  \end{bmatrix},
  \label{h_hat_g}
\end{align}
where the $2\times 2$ matrix, the second term on the right side of \eqref{h_hat_g}, is a nilpotent matrix that becomes a zero matrix when squared.
We call $\bm{\mathcal{\Hat{H}}}$ the Hamiltonian in this paper. 
By using the Hamiltonian, the fermion-type fundamental equation is defined as 
\begin{align}
  {\rm i} \, \frac{{\rm d}}{{\rm d}t}  \, \bm{\Hat{x}}(t) = \bm{\mathcal{\Hat{H}}} \, \bm{\Hat{x}}(t). 
 \label{fundamental_equation_eom_3}
\end{align}

Since the fermion-type fundamental equation (\ref{fundamental_equation_eom_3}) is a first-order differential equation with respect to time, it can describe causal relationships. 
Also, with regard to user dynamics, the solutions of the fermion-type fundamental equation (\ref{fundamental_equation_eom_3}) can generate solutions of the original wave equation (\ref{eom}). 
Details are as follows. 
From the fermion-type fundamental equation (\ref{fundamental_equation_eom_3}), we obtain the second derivative of $\bm{\Hat{x}}(t)$ with respect to time $t$ as follows:
\begin{align}
  \frac{{\rm d}^{2}\bm{\Hat{x}}(t)}{{\rm d}t^{2}}
  &=-\bm{\mathcal{\hat{H}}}^{2} \, \bm{\hat{x}}(t)\notag\\
  &= - \! \left\{ \! \left(\bm{\mathcal{D}}
  \!-\! \frac{1}{2}(\bm{\mathcal{A}}+(\sqrt{\bm{\mathcal{D}}})^{-1} \, \bm{\mathcal{A}} \, \sqrt{\bm{\mathcal{D}}})\right) \! \otimes \!
  \begin{bmatrix}
    1 & 0\\
    0 & 1
  \end{bmatrix}\right.
  \notag\\
  &\qquad\left.
  -\left(\bm{\mathcal{A}}-(\sqrt{\bm{\mathcal{D}}})^{-1} \, \bm{\mathcal{A}} \, \sqrt{\bm{\mathcal{D}}}\right)\otimes
  \frac{1}{2}
  \begin{bmatrix}
    0 & 1\\
    1 & 0
  \end{bmatrix}\right\} \, \bm{\Hat{x}}(t).
  \label{differential_two}
\end{align}
As a special case, when $\bm{\mathcal{A}}-(\sqrt{\bm{\mathcal{D}}})^{-1} \, \bm{\mathcal{A}} \, \sqrt{\bm{\mathcal{D}}} = \bm{O}$, i.e. weighted nodal degrees of all nodes are equal, $\bm{\mathcal{\Hat{H}}}$ is the square root matrix of the Laplacian matrix, that is, 
\[
\bm{\mathcal{\Hat{H}}}^2 = \bm{\mathcal{L}} \otimes \begin{bmatrix}1 & 0\\ 0 & 1 \end{bmatrix}.
\]
That is 
\[
  \frac{{\rm d}^{2}\bm{\Hat{x}}(t)}{{\rm d}t^{2}}
  =- \left(\left(\bm{\mathcal{D}}
  - \bm{\mathcal{A}}\right) \otimes 
  \begin{bmatrix}
    1 & 0\\
    0 & 1
  \end{bmatrix}\right)
  \bm{\Hat{x}}(t)
  =-
  \begin{bmatrix}
    \bm{\mathcal{L}} & 0\\
    0 & \bm{\mathcal{L}}
  \end{bmatrix}
  \bm{\Hat{x}}(t).
\]
In general cases, by extracting the differential equation for each of $\bm{x}^{+}(t)$ and $\bm{x}^{-}(t)$ from \eqref{differential_two}, we obtain
\begin{align}
  \frac{{\rm d}^{2}\bm{x}^{+}(t)}{{\rm d}t^{2}}
  &=-\left(\bm{\mathcal{D}}
  -\frac{1}{2} (\bm{\mathcal{A}}+(\sqrt{\bm{\mathcal{D}}})^{-1} \, \bm{\mathcal{A}} \sqrt{\bm{\mathcal{D}}})\right) \bm{x}^{+}(t)\notag\\
  &\quad\quad\quad
  +\frac{1}{2} \left(\bm{\mathcal{A}}-(\sqrt{\bm{\mathcal{D}}})^{-1} \, \bm{\mathcal{A}} \, \sqrt{\bm{\mathcal{D}}}\right) \, \bm{x}^{-}(t),
  \label{3.30}\\
  \frac{{\rm d}^{2}\bm{x}^{-}(t)}{{\rm d}t^{2}}
  &=-\left(\bm{\mathcal{D}}
  -\frac{1}{2} (\bm{\mathcal{A}}+(\sqrt{\bm{\mathcal{D}}})^{-1} \, \bm{\mathcal{A}} \sqrt{\bm{\mathcal{D}}})\right) \bm{x}^{-}(t)
  \notag\\
  &\quad\quad\quad
  +\frac{1}{2} \left(\bm{\mathcal{A}}-(\sqrt{\bm{\mathcal{D}}})^{-1} \, \bm{\mathcal{A}} \, \sqrt{\bm{\mathcal{D}}}\right) \, \bm{x}^{+}(t).
  \label{3.31}
\end{align}
Furthermore, by adding both sides of \eqref{3.30} and \eqref{3.31} respectively, we obtain
\begin{align}
  \frac{{\rm d}^{2}}{{\rm d}t^{2}} (\bm{x}^{+}(t) + \bm{x}^{-}(t)) &=
  - (\bm{\mathcal{D}} - \bm{\mathcal{A}}) \, (\bm{x}^{+}(t) + \bm{x}^{-}(t))
  \notag\\
  &=
  - \bm{\mathcal{L}} \, (\bm{x}^{+}(t) + \bm{x}^{-}(t)).
  \label{3.32}
\end{align}
This means that the sum of $\bm{x}^{+}(t)$ and $\bm{x}^{-}(t)$ are solutions of the original wave equation \eqref{eom}.

Finally, we give the Hamiltonian in the fermion-type fundamental equation (\ref{fundamental_equation_eom_3}) in a different form. 
The Hamiltonian (\ref{h_hat_g}) is equivalent to
\begin{align}
  \bm{\mathcal{\Hat{H}}}
  :=\bm{\mathcal{H}}\otimes \frac{1}{2}
  \begin{bmatrix}
    +1 & +1\\
    -1 & -1
  \end{bmatrix}
  +\sqrt{\bm{\mathcal{D}}}\otimes \frac{1}{2}
  \begin{bmatrix}
    +1 & -1\\
    +1 & -1
  \end{bmatrix}. 
  \label{Hamiltonian}
\end{align}
This is a convenient form for deriving solutions of the fundamental equation (\ref{fundamental_equation_eom_3}).

\section{Closed-Form Solution of the Fermion-Type Fundamental Equation (\ref{fundamental_equation_eom_3})}
\label{sec.4}
In this section, we derive a closed-form solution of the fermion-type fundamental equation (\ref{fundamental_equation_eom_3}) and compare it with the closed-form solutions of the boson-type fundamental equation (\ref{fundamental_equation_eom_two}).

\subsection{Preliminaries for the Algebraic Structure}
The solution of the fundamental equation (\ref{fundamental_equation_eom_3}) is formally expressed as 
\begin{align}
    \bm{\hat{x}}(t) &= \exp(-{\rm i} \,  \bm{\mathcal{\hat{H}}} \, t) \, \bm{\hat{x}}(0)
    \notag\\
    &= \sum_{k=0}^\infty \frac{1}{k!} \left(-{\rm i} \,  \bm{\mathcal{\hat{H}}} \, t\right)^k \, \bm{\hat{x}}(0) .
  \label{solution_formsl}
\end{align}
Before deriving the closed-form solution, we introduce important algebraic relations. 
Let us define
\[
\bm{\hat{a}} := \frac{1}{2} \,
\begin{bmatrix}
+1 & +1\\
-1 & -1
\end{bmatrix}, \ 
\bm{\hat{b}} := \frac{1}{2} \,
\begin{bmatrix}
+1 & -1\\
+1 & -1
\end{bmatrix}, \  
\bm{\hat{e}} :=
\begin{bmatrix}
+1 & 0\\
0 & +1
\end{bmatrix},
\]
where $\bm{\hat{a}}$ and $\bm{\hat{b}}$ appear in the Hamiltonian (\ref{Hamiltonian}). 
They satisfy the following anticommutation relations: 
\begin{align}
\{\bm{\hat{a}},\bm{\hat{b}}\} := \bm{\hat{a}}\bm{\hat{b}} + \bm{\hat{b}}\bm{\hat{a}} = \bm{\hat{e}}, \quad \bm{\hat{a}}^2 = \bm{\hat{b}}^2 = \bm{\mathrm{O}}\,\text{(null matrix)}. 
\end{align}
From the anticommutation relation of $\bm{\hat{a}}$ and $\bm{\hat{b}}$, we have 
\begin{align}
\bm{\hat{a}} \bm{\hat{b}} \bm{\hat{a}} &=(\bm{\hat{e}} - \bm{\hat{b}} \bm{\hat{a}})\, \bm{\hat{a}} =\bm{\hat{a}} - \bm{\hat{b}} \bm{\hat{a}}^2 = \bm{\hat{a}}, 
\notag\\
\bm{\hat{b}} \bm{\hat{a}} \bm{\hat{b}} &=(\bm{\hat{e}} - \bm{\hat{a}} \bm{\hat{b}}) \, \bm{\hat{b}} = \bm{\hat{b}} - \bm{\hat{a}} \bm{\hat{b}}^2 = \bm{\hat{b}}, 
\notag\\
\bm{\hat{a}} \bm{\hat{b}} \bm{\hat{a}} \bm{\hat{b}} &=(\bm{\hat{e}} -\bm{\hat{b}} \bm{\hat{a}}) \, \bm{\hat{a}} \bm{\hat{b}} = \bm{\hat{a}} \bm{\hat{b}}, 
\notag\\
\bm{\hat{b}} \bm{\hat{a}} \bm{\hat{b}} \bm{\hat{a}} &=(\bm{\hat{e}} - \bm{\hat{a}} \bm{\hat{b}}) \, \bm{\hat{b}} \bm{\hat{a}} = \bm{\hat{b}} \bm{\hat{a}} .
\notag
\end{align}
Therefore, when $\bm{\mathcal{\Hat{H}}}^k$ is expanded, the matrices that appear to the right of 
the Kronecker product are just 
\[
\bm{\hat{a}} \bm{\hat{b}} =
\frac{1}{2} \,
\begin{bmatrix}
+1 & -1\\
-1 & +1
\end{bmatrix}, \quad \text{and}\quad
\bm{\hat{b}} \bm{\hat{a}} =
\frac{1}{2} \,
\begin{bmatrix}
+1 & +1\\
+1 & +1
\end{bmatrix},
\]
in addition to $\bm{\hat{a}}$, $\bm{\hat{b}}$, and $\bm{\hat{e}}$.

\subsection{Closed-Form Solution of the Fundamental Equation (\ref{fundamental_equation_eom_3})}
By using $\bm{\mathcal{L}} = \sqrt{\bm{\mathcal{D}}} \, \bm{\mathcal{H}}$, 
the expansion of $\bm{\mathcal{\Hat{H}}}^n$ is expressed as 
\begin{align}
\bm{\mathcal{\Hat{H}}}^2 &= (\bm{\mathcal{H}} \otimes \bm{\hat{a}} + \sqrt{\bm{\mathcal{D}}} \otimes \bm{\hat{b}})\,( \bm{\mathcal{H}} \otimes \bm{\hat{a}} + \sqrt{\bm{\mathcal{D}}} \otimes \bm{\hat{b}})
\notag\\
&= \bm{\mathcal{H}} \, \sqrt{\bm{\mathcal{D}}} \otimes \bm{\hat{a}} \bm{\hat{b}}+ \bm{\mathcal{L}} \otimes \bm{\hat{b}} \bm{\hat{a}}, 
\notag\\
\bm{\mathcal{\Hat{H}}}^3 &= ( \bm{\mathcal{H}} \, \sqrt{\bm{\mathcal{D}}} \otimes \bm{\hat{a}} \bm{\hat{b}}+  \bm{\mathcal{L}} \otimes \bm{\hat{b}} \bm{\hat{a}})\,( \bm{\mathcal{H}} \otimes \bm{\hat{a}} +  \sqrt{\bm{\mathcal{D}}} \otimes\bm{\hat{b}})
\notag\\
&= \bm{\mathcal{H}} \,\bm{\mathcal{L}} \otimes \bm{\hat{a}} +  \bm{\mathcal{L}} \,\sqrt{\bm{\mathcal{D}}} \otimes \bm{\hat{b}},
\notag\\
\bm{\mathcal{\Hat{H}}}^4 &= (\bm{\mathcal{H}} \,\bm{\mathcal{L}} \otimes \bm{\hat{a}} +  \bm{\mathcal{L}} \, \sqrt{\bm{\mathcal{D}}} \otimes \bm{\hat{b}})\,( \bm{\mathcal{H}} \otimes \bm{\hat{a}} +  \sqrt{\bm{\mathcal{D}}} \otimes \bm{\hat{b}})
\notag\\
&=  \bm{\mathcal{H}} \, \bm{\mathcal{L}} \, \sqrt{\bm{\mathcal{D}}} \otimes \bm{\hat{a}}\bm{\hat{b}}+ \bm{\mathcal{L}}^2 \otimes \bm{\hat{b}}\bm{\hat{a}},
\notag\\
\bm{\mathcal{\Hat{H}}}^5 &= ( \bm{\mathcal{H}} \, \bm{\mathcal{L}} \, \sqrt{\bm{\mathcal{D}}} \otimes \bm{\hat{a}} \bm{\hat{b}} + \bm{\mathcal{L}}^2 \otimes \bm{\hat{b}}\bm{\hat{a}})\,( \bm{\mathcal{H}} \otimes \bm{\hat{a}} +  \sqrt{\bm{\mathcal{D}}} \otimes \bm{\hat{b}})
\notag\\
&= \bm{\mathcal{H}} \, \bm{\mathcal{L}}^2 \otimes \bm{\hat{a}} +   \bm{\mathcal{L}}^2 \, \sqrt{\bm{\mathcal{D}}} \otimes \bm{\hat{b}},
\notag\\
\bm{\mathcal{\Hat{H}}}^6 &= \bm{\mathcal{H}} \, \bm{\mathcal{L}}^2 \, \sqrt{\bm{\mathcal{D}}} \otimes \bm{\hat{a}} \bm{\hat{b}} + \bm{\mathcal{L}}^3 \otimes \bm{\hat{b}} \bm{\hat{a}},
\notag\\
\bm{\mathcal{\Hat{H}}}^7 &= \bm{\mathcal{H}} \, \bm{\mathcal{L}}^3 \otimes \bm{\hat{a}}+   \bm{\mathcal{L}}^3 \, \sqrt{\bm{\mathcal{D}}} \otimes \bm{\hat{b}}. 
\notag
\end{align}
In general, by using $\bm{\mathcal{H}} = (\sqrt{\bm{\mathcal{D}}})^{-1}\,\bm{\mathcal{L}}$, we easily have 
\begin{align}
\bm{\mathcal{\Hat{H}}}^{2k} &= (\sqrt{\bm{\mathcal{D}}})^{-1} \, \bm{\mathcal{L}}^k \, \sqrt{\bm{\mathcal{D}}}\otimes \bm{\hat{a}}\bm{\hat{b}}+ \bm{\mathcal{L}}^k \otimes \bm{\hat{b}} \bm{\hat{a}}, 
\label{H^even}\\
\bm{\mathcal{\Hat{H}}}^{2k+1} &= \bm{\mathcal{H}} \, \bm{\mathcal{L}}^k \otimes \bm{\hat{a}} +   \bm{\mathcal{L}}^k \, \sqrt{\bm{\mathcal{D}}} \otimes \bm{\hat{b}}, 
\label{H^odd}
\end{align}
for $k \ge 0$. 
Using these relations, we  can express the solution (\ref{solution_formsl}) of the fundamental equation~(\ref{fundamental_equation_eom_3}) in closed-form. 
We substitute (\ref{H^odd}) and (\ref{H^even}) into 
\begin{align}
\bm{\Hat{x}}(t) &= \exp[-\ii\bm{\mathcal{\Hat{H}}}\,t] \, \bm{\Hat{x}}(0) \notag\\
&= (\cos(\bm{\mathcal{\Hat{H}}}\,t) -\ii\,\sin(\bm{\mathcal{\Hat{H}}}\,t))\, \bm{\Hat{x}}(0)
\notag\\
&= \left(\sum_{k=0}^\infty (-1)^k \,\frac{\bm{(\mathcal{\Hat{H}}}\,t)^{2k}}{(2k)!} 
- \ii \, \sum_{k=0}^\infty (-1)^k \, \frac{\bm{(\mathcal{\Hat{H}}}\,t)^{2k+1}}{(2k+1)!}\right) \bm{\Hat{x}}(0). 
\notag
\end{align}
Both infinite sums are expressed as trigonometric functions as
\begin{align*}
  &\cos(\bm{\mathcal{\Hat{H}}}\,t) = \sum_{k=0}^\infty (-1)^k \, \frac{\bm{(\mathcal{\Hat{H}}}\,t)^{2k}}{(2k)!} 
  \notag\\
  &=(\sqrt{\bm{\mathcal{D}}})^{-1} \, \left(\bm{I} -  \bm{\mathcal{L}}\,\frac{t^{2}}{2!}  + \bm{\mathcal{L}}^{2} \,\frac{t^{4}}{4!} - \cdots \right)  \sqrt{\bm{\mathcal{D}}}\otimes \bm{\hat{a}}\bm{\hat{b}}
  \notag\\
  &\quad\quad\quad+\left(\bm{I}-\bm{\mathcal{L}}\,\frac{t^{2}}{2!} \, +\bm{\mathcal{L}}^{2}\,\frac{t^{4}}{4!} -\cdots \right)\otimes \bm{\hat{b}}\bm{\hat{a}}
  \notag\\
  &= (\sqrt{\bm{\mathcal{D}}})^{-1} \bm{P} \left(\bm{I} - \bm{\Omega}^{2}\,\frac{t^{2}}{2!}  +  \bm{\Omega}^{4}\,\frac{t^{4}}{4!} - \cdots \right) \bm{P}^{-1} \sqrt{\bm{\mathcal{D}}}\otimes \bm{\hat{a}}\bm{\hat{b}}
  \notag\\
  &\quad\quad+\bm{P} \, \left(\bm{I}-\bm{\Omega}^{2}\,\frac{t^{2}}{2!} \, +\bm{\Omega}^{4}\,\frac{t^{4}}{4!}-\cdots \right) \, \bm{P}^{-1}\otimes \bm{\hat{b}}\bm{\hat{a}}
  \notag\\
  &=
  (\sqrt{\bm{\mathcal{D}}})^{-1} \, \bm{P}\,
  \cos(\bm{\Omega}\,t) \,
  \bm{P}^{-1} \, \sqrt{\bm{\mathcal{D}}} \otimes \bm{\hat{a}}\bm{\hat{b}}
  \notag\\
  &\quad\quad
  +\bm{P} \, \cos(\bm{\Omega}\,t) \, \bm{P}^{- 1} \otimes \bm{\hat{b}}\bm{\hat{a}},
\end{align*}
and
\begin{align*}
  &\sin(\bm{\mathcal{\Hat{H}}}\,t) = \sum_{k=0}^\infty (-1)^k \, \frac{\bm{(\mathcal{\Hat{H}}}\,t)^{2k+1}}{(2k+1)!}
  \notag\\
  &= \left(\bm{\mathcal{H}}\,t - \bm{\mathcal{H}} \, \bm{\mathcal{L}}\,\frac{t^{3}}{3!} +\bm{\mathcal{H}} \, \bm{\mathcal{L}}^{2}\,\frac{t^{5}}{5!} -\cdots \right)\otimes \bm{\hat{a}}
  \notag\\
  &\quad + \left(\sqrt{\bm{\mathcal{D}}}\,t - \bm{\mathcal{L}} \, \sqrt{\bm{\mathcal{D}}}\,\frac{t^{3}}{3!} + \bm{\mathcal{L}}^{2} \, \sqrt{\bm{\mathcal{D}}}\,\frac{t^{5}}{5!} - \cdots  \right) \otimes \bm{\hat{b}}
  \notag\\
  &= (\sqrt{\bm{\mathcal{D}}})^{-1} \bm{P} \bm{\Omega}\left(\bm{\Omega}\,t - \bm{\Omega}^{3}\,\frac{t^{3}}{3!}  + \bm{\Omega}^{5} \, \frac{t^{5}}{5!} - \cdots  \right) \bm{P}^{-1} \otimes \bm{\hat{a}}
  \notag\\
  &\quad + \bm{P} \bm{\mho} \left(\bm{\Omega}\,t - \bm{\Omega}^{3}\,\frac{t^{3}}{3!} + \bm{\Omega}^{5} \, \frac{t^{5}}{5!} - \cdots \right) \bm{P}^{-1} \sqrt{\bm{\mathcal{D}}} \otimes \bm{\hat{b}}
  \notag\\
  &=
  (\sqrt{\bm{\mathcal{D}}})^{-1} \, \bm{P} \, \bm{\Omega} \, \sin(\bm{\Omega}\,t) \, \bm{P}^{-1}\otimes \bm{\hat{a}}
  \notag\\
  &\quad\quad
  +\bm{P} \,
  \bm{\mho} \, \sin(\bm{\Omega}\,t) \, \bm{P}^{-1} \, \sqrt{\bm{\mathcal{D}}}\otimes \bm{\hat{b}}, 
\end{align*}
where 
\begin{align*}
\cos(\bm{\Omega}\,t) &= \mathrm{diag}(\cos(\omega_0\,t),\,\cos(\omega_1\,t),\,\dots,\,\cos(\omega_{n-1}\,t)),
\\
\sin(\bm{\Omega}\,t) &= \mathrm{diag}(\sin(\omega_0\,t),\,\sin(\omega_1\,t),\,\dots,\,\sin(\omega_{n-1}\,t)),
\\
\bm{\mho} &:= \text{diag}(0,\,1/\omega_1,\,\dots,\,1/\omega_{n-1}). 
\end{align*}
Therefore, the closed-form solution is given by
\begin{align}
\bm{\Hat{x}}(t) 
&= \Big(\big(\sqrt{\bm{\mathcal{D}}}\big)^{-1} \, \bm{P} \, \cos(\bm{\Omega}\,t) \, \bm{P}^{-1} \, \sqrt{\bm{\mathcal{D}}} \otimes \bm{\hat{a}}\bm{\hat{b}}
\notag\\
&\qquad + \bm{P} \, \cos(\bm{\Omega}\,t) \, \bm{P}^{-1} \otimes \bm{\hat{b}}\bm{\hat{a}}\Big) \, \bm{\Hat{x}}(0) 
\notag\\
&\quad -\ii \, \Big(\big(\sqrt{\bm{\mathcal{D}}}\big)^{-1} \, \bm{P} \, \bm{\Omega}\,\sin(\bm{\Omega}\,t) \, \bm{P}^{-1} \otimes \bm{\hat{a}}
\notag\\
&\qquad + \bm{P} \,  \bm{\mho} \, \sin(\bm{\Omega}\,t) \, \bm{P}^{-1} \, \sqrt{\bm{\mathcal{D}}} \otimes \bm{\hat{b}} \Big) \, \bm{\Hat{x}}(0).
\label{eq:sol-F}
\end{align}

\subsection{Comparison of the Solutions of Two Types of Fundamental Equations (\ref{fundamental_equation_eom_two}) and (\ref{fundamental_equation_eom_3})}
The solution (\ref{eq:sol-F}) of the fermion-type fundamental equation (\ref{fundamental_equation_eom_3}) looks very different in form from the solution (\ref{solution-sqrt{L}}) of the boson-type fundamental equation (\ref{fundamental_equation_eom_two}).
Let us confirm if they are indeed different. 
To distinguish between the two we denote the solution (\ref{eq:sol-F}) as $\bm{\Hat{x}}_\mathrm{f}(t)$ and the solution (\ref{solution-sqrt{L}}) as $\bm{\Hat{x}}_\mathrm{b}(t)$.

As a simple example, let us consider a regular graph, that is, all nodes have degree of $d$, the same value. 
For the regular graph, we obtain 
\[
\sqrt{\bm{\mathcal{D}}} = \sqrt{d} \, \bm{I}; 
\]
it is commutable with any $n\times n$ matrix. 
Therefore, the solution (\ref{eq:sol-F}) becomes
\begin{align}
\bm{\Hat{x}}_\mathrm{f}(t) 
&= \Big(\bm{P} \, \cos(\bm{\Omega}\,t) \, \bm{P}^{-1} \otimes (\bm{\hat{a}}\bm{\hat{b}} + \bm{\hat{b}}\bm{\hat{a}})\Big) \, \bm{\Hat{x}}(0)
\notag\\
&\quad -\ii \, \Big(\frac{1}{\sqrt{d}}\, \bm{P} \, \bm{\Omega}\,\sin(\bm{\Omega}\,t) \, \bm{P}^{-1} \otimes \bm{\hat{a}}
\notag\\
&\qquad + \sqrt{d} \, \bm{P} \,  \bm{\mho} \, \sin(\bm{\Omega}\,t) \, \bm{P}^{-1} \otimes \bm{\hat{b}} \Big) \, \bm{\Hat{x}}(0) 
\notag\\
&= \Big(\bm{P} \, \cos(\bm{\Omega}\,t) \, \bm{P}^{-1} \otimes \bm{\hat{e}}\Big) \, \bm{\Hat{x}}(0)
\notag\\
&\quad -\ii \, \Big(\frac{1}{\sqrt{d}}\,\bm{P} \, \bm{\Omega}\,\sin(\bm{\Omega}\,t) \, \bm{P}^{-1} \otimes \bm{\hat{a}}
\notag\\
&\qquad + \sqrt{d} \, \bm{P} \,  \bm{\mho} \, \sin(\bm{\Omega}\,t) \, \bm{P}^{-1} \otimes \bm{\hat{b}} \Big) \, \bm{\Hat{x}}(0) 
,
\label{x_f}
\end{align}
for the initial condition $\bm{\Hat{x}}_\mathrm{f}(0) = \bm{\Hat{x}}(0)$. On the other hand, solution (\ref{solution-sqrt{L}}) is expressed as 
\begin{align}
\bm{\Hat{x}}_\mathrm{b}(t) &= \Big(\bm{P} \, \cos(\bm{\Omega}\,t) \, \bm{P}^{-1} \otimes \bm{\hat{e}}\Big)\, \bm{\Hat{x}}(0)
\notag\\
&-\ii \, \Big(\bm{P} \, \sin(\bm{\Omega}\,t) \, \bm{P}^{-1} \otimes 
\begin{bmatrix}
+1 & 0\\
0 & 0
\end{bmatrix}
\notag\\
&\quad\quad\quad - \bm{P} \, \sin(\bm{\Omega}\,t) \, \bm{P}^{-1} \otimes 
\begin{bmatrix}
0 & 0\\
0 & +1
\end{bmatrix}
\Big)\, \bm{\Hat{x}}(0), 
\label{x_b}
\end{align}
for the initial condition $\bm{\Hat{x}}_\mathrm{b}(0) = \bm{\Hat{x}}(0)$.

A comparison of (\ref{x_f}) and (\ref{x_b}) shows they are different although we can adjust that the real parts to be the same. 
Therefore, $\bm{\Hat{x}}_\mathrm{f}(t)$ and $\bm{\Hat{x}}_\mathrm{b}(t)$ express different solutions. 

\section{General Solutions of the Wave Equation Derived from Fundamental Equations (\ref{fundamental_equation_eom_two}) and (\ref{fundamental_equation_eom_3})}
\label{sec.5}
In this section, we derive the solutions of the original wave equation \eqref{eom} from the closed-form solutions of the two different fundamental equations (\ref{fundamental_equation_eom_two}) and (\ref{fundamental_equation_eom_3}), and verify that they follow the original wave equation.

\subsection{General Solutions of the Wave Equation}
As shown in Sec.~\ref{sec.3}, solutions of the fundamental equations (\ref{fundamental_equation_eom_two}) and (\ref{fundamental_equation_eom_3}) should be able to generate general solutions of the original wave equation \eqref{eom}.
That is, for solutions $\bm{x}^+(t)$ and $\bm{x}^-(t)$ of the fundamental equations (\ref{fundamental_equation_eom_two}) and (\ref{fundamental_equation_eom_3}), the sum of them, $\bm{x}(t)=\bm{x}^+(t)+\bm{x}^-(t)$, should be able to give a general solution. 

Let us consider the boson-type fundamental equation (\ref{fundamental_equation_eom_two}) and denote its solution as 
\[
\bm{\hat{x}}_\mathrm{b}(t) = \bm{x}_\mathrm{b}^+(t) \otimes 
\begin{pmatrix}
1\\
0
\end{pmatrix}
+\bm{x}_\mathrm{b}^-(t) \otimes 
\begin{pmatrix}
0\\
1
\end{pmatrix}.
\] 

From the solutions (\ref{solution-x-pm}) of the boson-type fundamental equation (\ref{fundamental_equation_eom_two}), $\bm{x}_\mathrm{b}(t):=\bm{x}_\mathrm{b}^+(t)+\bm{x}_\mathrm{b}^-(t)$ is expressed as 
\begin{align}
\bm{x}_\mathrm{b}(t) 
&= \bm{P} \, \exp(-\ii\,\bm{\Omega}\,t) \, \bm{P}^{-1} \, \bm{x}^+(0) 
\notag\\
&\quad {}+ \bm{P} \, \exp(+\ii\,\bm{\Omega}\,t) \, \bm{P}^{-1} \, \bm{x}^-(0), 
\label{solutionx_b}
\end{align}
for the initial condition $\bm{x}_\mathrm{b}(0)= \bm{x}^+(0) + \bm{x}^-(0)$. 
The second-order differential of $\bm{x}_\mathrm{b}(t)$ with respect to $t$ gives
\begin{align}
\frac{\dd^2}{\dd t^2} \, \bm{x}_\mathrm{b}(t) 
&= \frac{\dd^2}{\dd t^2} \, \bm{P} \, \exp(-\ii\,\bm{\Omega}\,t) \, \bm{P}^{-1} \, \bm{x}^+(0) 
\notag\\
&\quad {}+ \frac{\dd^2}{\dd t^2} \, \bm{P} \, \exp(+\ii\,\bm{\Omega}\,t) \, \bm{P}^{-1} \, \bm{x}^-(0)
\notag\\
&= -\ii\,\frac{\dd}{\dd t} \, \bm{P} \,\bm{\Omega}\, \exp(-\ii\,\bm{\Omega}\,t) \, \bm{P}^{-1} \, \bm{x}^+(0) 
\notag\\
&\quad {}+\ii\, \frac{\dd}{\dd t} \, \bm{P} \,\bm{\Omega}\, \exp(+\ii\,\bm{\Omega}\,t) \, \bm{P}^{-1} \, \bm{x}^-(0)
\notag\\
&= -\bm{P}\, \bm{\Omega}^2\, \exp(-\ii\,\bm{\Omega}\,t) \, \bm{P}^{-1} \, \bm{x}^+(0) 
\notag\\
&\quad {}-\bm{P} \,\bm{\Omega}^2\, \exp(+\ii\,\bm{\Omega}\,t) \, \bm{P}^{-1} \, \bm{x}^-(0)
\notag\\
&= -\bm{P}\,\bm{\Omega}^2\,\bm{P}^{-1} \, \bm{P} \,\exp(-\ii\,\bm{\Omega}\,t) \, \bm{P}^{-1} \, \bm{x}^+(0) 
\notag\\
&\quad {}-\bm{P} \,\bm{\Omega}^2\,\bm{P}^{-1} \, \bm{P}\, \exp(+\ii\,\bm{\Omega}\,t) \, \bm{P}^{-1} \, \bm{x}^-(0)
\notag\\
&= -\bm{\mathcal{L}} \, \bm{x}_\mathrm{b}(t). 
\end{align}
This means $\bm{x}_\mathrm{b}(t)$ is a solution of the original wave equation (\ref{eom}). 
For comparison, we rewrite (\ref{solutionx_b}) in the following form: 
\begin{align}
\bm{x}_\mathrm{b}(t) &= \bm{P} \, \cos(\bm{\Omega}\,t) \, \bm{P}^{-1} \, (\bm{x}^+(0)+\bm{x}^-(0))
\notag\\
&\quad {}-\ii \,\bm{P} \,  \sin(\bm{\Omega}\,t) \, \bm{P}^{-1} \,(\bm{x}^+(0)-\bm{x}^-(0)),
\label{solutionx_b2}
\end{align}

Next, we consider the fermion-type fundamental equation (\ref{fundamental_equation_eom_3}) and denote its solution as 
\[
\bm{\hat{x}}_\mathrm{f}(t) = \bm{x}_\mathrm{f}^+(t) \otimes 
\begin{pmatrix}
1\\
0
\end{pmatrix}
+\bm{x}_\mathrm{f}^-(t) \otimes 
\begin{pmatrix}
0\\
1
\end{pmatrix}.
\] 
Solution (\ref{eq:sol-F}) can be rewritten as separate entities for $\bm{x}_\mathrm{f}^+(t)$ and $\bm{x}_\mathrm{f}^-(t)$ as follows: 
\begin{align}
\bm{x}_\mathrm{f}^+(t) &= \big(\sqrt{\bm{\mathcal{D}}}\big)^{-1} \, \bm{P} \, \cos(\bm{\Omega}\,t) \, \bm{P}^{-1} \, \sqrt{\bm{\mathcal{D}}}\,\,\frac{\bm{x}^+(0)-\bm{x}^-(0)}{2}
\notag\\
&\quad {}+ \bm{P} \, \cos(\bm{\Omega}\,t) \, \bm{P}^{-1} \,\frac{\bm{x}^+(0)+\bm{x}^-(0)}{2}
\notag\\
&\quad {}-\ii \,\big(\sqrt{\bm{\mathcal{D}}}\big)^{-1} \, \bm{P} \, \bm{\Omega}\,\sin(\bm{\Omega}\,t) \, \bm{P}^{-1} \,\frac{\bm{x}^+(0)+\bm{x}^-(0)}{2}
\notag\\
&\quad {}-\ii \,\bm{P} \,  \bm{\mho} \, \sin(\bm{\Omega}\,t) \, \bm{P}^{-1} \sqrt{\bm{\mathcal{D}}} \,\,\frac{\bm{x}^+(0)-\bm{x}^-(0)}{2},
\\
\bm{x}_\mathrm{f}^-(t) &= -\big(\sqrt{\bm{\mathcal{D}}}\big)^{-1} \, \bm{P} \, \cos(\bm{\Omega}\,t) \, \bm{P}^{-1} \, \sqrt{\bm{\mathcal{D}}}\,\,\frac{\bm{x}^+(0)-\bm{x}^-(0)}{2}
\notag\\
&\quad {}+ \bm{P} \, \cos(\bm{\Omega}\,t) \, \bm{P}^{-1} \,\frac{\bm{x}^+(0)+\bm{x}^-(0)}{2}
\notag\\
&\quad {}+\ii \,\big(\sqrt{\bm{\mathcal{D}}}\big)^{-1} \, \bm{P} \, \bm{\Omega}\,\sin(\bm{\Omega}\,t) \, \bm{P}^{-1} \,\frac{\bm{x}^+(0)+\bm{x}^-(0)}{2}
\notag\\
&\quad {}-\ii \,\bm{P} \,  \bm{\mho} \, \sin(\bm{\Omega}\,t) \, \bm{P}^{-1} \sqrt{\bm{\mathcal{D}}} \,\,\frac{\bm{x}^+(0)-\bm{x}^-(0)}{2},
\end{align}
for the initial condition $\bm{x}_\mathrm{f}(0)= \bm{x}^+(0) + \bm{x}^-(0)$. 
The sum $\bm{x}_\mathrm{f}(t) := \bm{x}_\mathrm{f}^+(t) + \bm{x}_\mathrm{f}^-(t)$ gives 
\begin{align}
\bm{x}_\mathrm{f}(t) &= \bm{P} \, \cos(\bm{\Omega}\,t) \, \bm{P}^{-1} \, (\bm{x}^+(0)+\bm{x}^-(0))
\notag\\
&\quad {}-\ii \,\bm{P} \,  \bm{\mho} \, \sin(\bm{\Omega}\,t) \, \bm{P}^{-1} \sqrt{\bm{\mathcal{D}}} \,(\bm{x}^+(0)-\bm{x}^-(0)).
\label{x_f(0)}
\end{align}

The second-order differential of $\bm{x}_\mathrm{f}(t)$ with respect to $t$ gives
\begin{align}
\frac{\dd^2}{\dd t^2} \, \bm{x}_\mathrm{f}(t) 
&= \frac{\dd^2}{\dd t^2} \, \bm{P} \, \cos(\bm{\Omega}\,t) \, \bm{P}^{-1} \,(\bm{x}^+(0)+\bm{x}^-(0))
\notag\\
&\quad {}-\ii \, \frac{\dd^2}{\dd t^2} \, \bm{P} \,  \bm{\mho} \, \sin(\bm{\Omega}\,t) \, \bm{P}^{-1} \sqrt{\bm{\mathcal{D}}} \,(\bm{x}^+(0)-\bm{x}^-(0))
\notag\\
&= -\frac{\dd}{\dd t} \, \bm{P} \,\bm{\Omega}\, \sin(\bm{\Omega}\,t) \, \bm{P}^{-1} \,(\bm{x}^+(0)+\bm{x}^-(0))
\notag\\
&\quad {}-\ii \, \frac{\dd}{\dd t} \, \bm{P} \, \bm{\mho} \, \bm{\Omega} \, \cos(\bm{\Omega}\,t) \, \bm{P}^{-1} \sqrt{\bm{\mathcal{D}}} \,(\bm{x}^+(0)-\bm{x}^-(0))
\notag\\
&= -\bm{P} \,\bm{\Omega}^2\, \cos(\bm{\Omega}\,t) \, \bm{P}^{-1} \,(\bm{x}^+(0)+\bm{x}^-(0))
\notag\\
&\quad {}+\ii \, \bm{P} \,\bm{\Omega}\,\sin(\bm{\Omega}\,t) \, \bm{P}^{-1} \sqrt{\bm{\mathcal{D}}} \,(\bm{x}^+(0)-\bm{x}^-(0))
\notag\\
&= -\bm{P} \,\bm{\Omega}^2\,\bm{P}^{-1}\, \bm{P}\cos(\bm{\Omega}\,t) \, \bm{P}^{-1} \,(\bm{x}^+(0)+\bm{x}^-(0))
\notag\\
&\quad {}- \bm{P} \,\bm{\Omega}^2\,\bm{P}^{-1}\,(-\ii \, \bm{P}\,\bm{\mho}\,\sin(\bm{\Omega}\,t) \, \bm{P}^{-1}) \sqrt{\bm{\mathcal{D}}} 
\notag\\
& \qquad \qquad \qquad \qquad \qquad \times (\bm{x}^+(0)-\bm{x}^-(0))
\notag\\
&= -\bm{\mathcal{L}} \, \bm{x}_\mathrm{f}(t). 
\end{align}
This means $\bm{x}_\mathrm{f}(t)$ is also a solution of the original wave equation (\ref{eom}). 

The solution $\bm{x}_\mathrm{b}(t)$ of (\ref{solutionx_b}) (or equivalently (\ref{solutionx_b2})) is obtained from the boson-type fundamental equation (\ref{fundamental_equation_eom_two}), and it is also derived directly by solving the original wave equation (\ref{eom}). 
In addition, since the solution (\ref{solutionx_b}) includes two arbitrary initial conditions, $\bm{x}^+(0)$ and $\bm{x}^-(0)$, it is a general solution of the original wave equation (\ref{eom}).
On the other hand, the solution $\bm{x}_\mathrm{f}(t)$ of (\ref{x_f(0)}) is also a solution of the original wave equation (\ref{eom}) although it is not derived directly by solving the original wave equation (\ref{eom}). 
The solution $\bm{x}_\mathrm{f}(t)$ is different, at least in form, from the well-known solution \eqref{solutionx_b2}. 
However, since the solution (\ref{x_f(0)}) also includes two arbitrary initial conditions, $\bm{x}^+(0)$ and $\bm{x}^-(0)$, it is a general solution of the original wave equation (\ref{eom}). 
That is, both solutions (\ref{solutionx_b2}) and (\ref{x_f(0)}) give the same set of solutions, although the actual solutions for the given initial condition are different. 
Such difference implies the importance of the fermion-type fundamental equation (\ref{fundamental_equation_eom_3}) because it is more suitable for describing user dynamics in OSNs. 
We can expect that the user dynamics described by the fermion-type fundamental equation (\ref{fundamental_equation_eom_3}) include unknown characteristics.

\section{Conclusion}
\label{sec.6}
This paper has derived the closed-form solution of the fermion-type fundamental equation, which is different from the closed-form solution of the boson-type fundamental equation.
In addition, we have derived a closed-form solution of the original wave equation from the solution of the fermion-type fundamental equation. 
The solution of the original wave equation derived from the fermion-type fundamental equation looks in a different form compared with the well-known general solution of the original wave equation directly obtained from the boson-type fundamental equation.

Both solutions derived from two fundamental equations give general solutions of the original wave equation, although the actual solutions for the given initial condition are different.
This fact implies the significance of the fermion-type fundamental equation. 
In the future, we will investigate the structure of solutions of the fermion-type fundamental equation.

\section*{Acknowledgment}
This research was supported by Grant-in-Aid for Scientific Research (B) No.~19H04096 (2019--2021) and No.~20H04179 (2020--2022) from the Japan Society for the Promotion of Science (JSPS).

\newpage



\end{document}